\documentclass[letterpaper, 10 pt, conference]{ieeeconf}
\IEEEoverridecommandlockouts
\usepackage{cite}
\usepackage{amsmath,amssymb,amsfonts}
\usepackage[english]{babel}
\usepackage{algorithmic}
\usepackage{graphicx}
\usepackage{textcomp}
\usepackage{optidef}
\usepackage{listings}
\usepackage[ruled,vlined]{algorithm2e}
\newtheorem{theorem}{Theorem}
\newtheorem{corollary}{Corollary}
\newtheorem{assumption}{Assumption}

\newtheorem{remark}{Remark}

\newtheorem{defn}{Definition}
\newtheorem{ex}{Example}

\usepackage{caption}
\usepackage{subcaption}

\usepackage{placeins}
\usepackage{mathtools}
\usepackage{lipsum}
\usepackage{braket}

\usepackage{bbding}
\usepackage{pifont}
\usepackage{wasysym}
\usepackage{amssymb}
\usepackage{multirow}
\usepackage{xcolor}

\newcommand{\revres}{}
\newcommand{\reviewer}[1]{}

\newcommand{\Rspace}{\mathbb{R}}

\newcommand{\id}{\textbf{I}}

\newcommand{\Ss}{\mathcal{S}}
\newcommand{\B}{\mathcal{B}}
\newcommand{\X}{\mathcal{X}}

\newcommand{\Z}{\mathcal{Z}}
\newcommand{\U}{\mathcal{U}}
\newcommand{\R}{\mathcal{R}}

\newcommand{\C}{\mathcal{C}}

\newcommand{\FSUSOL}{\Psi}
\newcommand{\FSUSCL}{\Phi}

\newcommand{\suc}{\text{Suc}}

\newcommand{\domfFX}{\text{D}(\FSUSCL)}
\newcommand{\domfFXU}{\text{D}(\FSUSOL)}
\newcommand{\SIM}{\Theta}
\newcommand{\domSIM}{\text{D}(\SIM)}

\usepackage{soul}

\def\BibTeX{{\rm B\kern-.05em{\sc i\kern-.025em b}\kern-.08em
    T\kern-.1667em\lower.7ex\hbox{E}\kern-.125emX}}
\begin{document}
\title{Successor Sets of Discrete-time Nonlinear Systems Using Hybrid Zonotopes}
\author{Jacob A. Siefert, Trevor J. Bird, Justin P. Koeln, Neera Jain, and Herschel C. Pangborn
\thanks{$*$The first two authors contributed equally to this work.}
\thanks{Jacob A. Siefert and Herschel C. Pangborn are with the Department of Mechanical Engineering, Pennsylvania State University, University Park, PA 16802 USA (e-mail: jas7031@psu.edu; hcpangborn@psu.edu).}
\thanks{Trevor J. Bird and Neera Jain are with the School of Mechanical Engineering, Purdue University, West Lafayette, IN 47907 USA (e-mail: bird6@purdue.edu; neerajain@purdue.edu).}
\thanks{Justin P. Koeln is with the  Mechanical Engineering Department, University of Texas at Dallas, Richardson, TX 75080-3021 USA (e-mail: justin.koeln@utdallas.edu).}
}

\maketitle
\thispagestyle{empty} 

\begin{abstract}
This paper presents identities for calculating over-approximated successor sets of discrete-time nonlinear systems using hybrid zonotopes. The proposed technique extends the state-update set construct, previously developed for linear hybrid systems, to nonlinear systems. Forward reachability of nonlinear systems can then be performed using only projection, intersection, and Cartesian product set operations with the state-update set. It is shown that use of an over-approximation of the state-update set yields over-approximations of successor sets. A technique to over-approximate a nonlinear function using a special ordered set approximation, equivalently represented as a hybrid zonotope, is then presented. A numerical example of a nonlinear system controlled by a piecewise-affine control law demonstrates that the approach provides a computationally-efficient and tight over-approximation of the closed-loop reachable set. 
\end{abstract}
\section{Introduction}
\label{sect:introduction}

Reachable sets are used to evaluate system performance and ensure constraint satisfaction in safety-critical applications.
For discrete-time systems, reachable sets are calculated by recursion of successor sets, also referred to as one-step forward reachable sets \cite{borrelli_predictive_2017}. For nonlinear systems and hybrid systems with continuous and discrete dynamics, the complexity and nonconvexity of successor sets can limit the scalability of existing approaches.

\subsubsection{Gaps in Literature}  Several methods have been proposed to calculate reachable sets of continuous-time nonlinear systems, including Hamilton-Jacobi reachability \cite{Bansal2017HamiltonJacobiRA,Bui2021RealTimeHR,HJreachDecomp2018,FastHJreachApprox2016}, optimization techniques \cite{OPTNL_mitchell2005time,OPTNL_chutinan2003computational}, monotonicity-based techniques \cite{MONO_angeli2003monotone,MONO_ramdani2008reachability}, and techniques based on abstractions of the state space \cite{Althoff_ReachSafetyAutonomousCars_2010,ConPolyZono_Althoff,SparsePolyZono_Althoff}. The latter of these closely mimic discrete-time reachability techniques in that they propagate reachable sets over time intervals and bound the effect of intersample dynamics. For this reason, many challenges that arise in calculating reachable sets of continuous-time systems using state-space abstraction methods also arise in discrete-time reachability. In many cases, abstraction of the state space is accomplished by approximating nonlinear functions with affine functions over partitioned regions of the domain, though higher-order abstractions are compatible with some set representations, e.g., \cite{ConPolyZono_Althoff,SparsePolyZono_Althoff}. State-space abstractions bound the error associated with higher-order terms, which when added to the affine abstractions, creates a collection of polyhedral over-approximations of the nonlinear function. State-space abstraction can either be done in a time-invariant \cite{ABS_INV_asarin2007hybridization,ABS_INV_asarin2003reachability} or time-varying manner \cite{ABS_TV_althoff2008reachability}. A primary challenge with time-invariant abstractions is related to their hybrid nature, i.e., reachable sets must be intersected with guards associated with the domain of each linearized region, which can cause exponential growth in the number of convex reachable sets. Unification methods exist to reduce the number of reachable sets, though this is computationally expensive \cite{Althoff_SetPropagationTechniques_2021}. Time-varying abstractions have been shown to enable tighter enclosures of reachable sets \cite{Althoff_SetPropagationTechniques_2021}, though algorithm performance is dependant on appropriate selection of tuning parameters. Reducing the need for manual tuning in time-varying state-space abstraction methods is an active area of research \cite{WETZLINGER_AdapReachNonLin,ABS_TV_wetzlinger2021adaptive}.

Similar to the methods for time-invariant abstractions of nonlinear continuous-time systems \cite{ABS_INV_asarin2007hybridization,ABS_INV_asarin2003reachability}, forward reachable sets of discrete-time linear hybrid systems 
may be determined using a collection of convex sets by partitioning the state space into locations separated by guards and applying techniques developed for linear systems within each location \cite{alur_algorithmic_1995,bemporad2003modeling}. Successive intersections with guards at each time step result in worst-case exponential growth in complexity, leading to computational-intractability for long time horizons. Recent work by the authors of this paper has addressed challenges in calculating forward reachable sets of discrete-time hybrid systems by defining a new construct called the \emph{state-update set}, which encodes all possible state transitions, and a new set representation called the \emph{hybrid zonotope}, which introduces binary factors into the set definition \cite{SiefertHybSUS,Bird_HybZono,Bird_HybZonoMPC,Bird_HybZonoUnionComp}. Hybrid zonotopes have been shown to enable scalable closed-form solutions of precursor and successor sets for broad classes of discrete-time linear hybrid systems using state-update sets \cite{SiefertHybSUS}. However, prior work has not explored applying hybrid zonotopes for reachability of more general nonlinear systems.

\subsubsection{Contribution} This paper provides closed-form identities for calculating over-approximated successor sets of discrete-time \emph{nonlinear} systems using hybrid zonotopes and state-update sets, building on previous results from \cite{SiefertHybSUS} that focused on linear hybrid systems. Using the proposed approach, over-approximations of successor sets can be computed with linear computational complexity with respect to the state dimension and linear memory complexity growth in time. In contrast to \cite{SiefertHybSUS}, this paper addresses open- and closed-loop dynamics separately and provides a  method to create a closed-loop state-update set by coupling an open-loop state-update set to a set-based representation of the control law, called the state-input map. Additionally, we show how special ordered set approximations of nonlinear functions can be represented as hybrid zonotopes, in turn enabling nonconvex over-approximations of nonlinear systems. 

\subsubsection{Outline} The remainder of this paper is organized as follows. Section \ref{sec:Prelims} provides notation and preliminary definitions. 
Section \ref{sect:SUS_reach} defines open-loop and closed-loop state-update sets and develops identities for successor sets. Section \ref{sec:HybZonoSUS} shows how to generate hybrid zonotopes equivalent to special ordered set approximations of nonlinear functions, which can be used to generate over-approximations of open-loop and closed-loop state-update sets. The numerical example in Section \ref{sect:Examples} demonstrates the creation of a closed-loop state-update set and calculation of forward reachable sets for a nonlinear system in closed-loop with a piecewise-affine controller.  Concluding remarks are made in Section \ref{sect:conclusion}.

\section{Preliminaries and Previous Work}
\label{sec:Prelims}

\subsubsection{Notation}
Matrices are denoted by uppercase letters, e.g., $G\in\Rspace^{n\times n_g}$, and sets by uppercase calligraphic letters, e.g., $\mathcal{Z}\subset\Rspace^{n}$. Vectors and scalars are denoted by lowercase letters. The $i^{th}$ column of a matrix $G$ is denoted $G_{(\cdot,i)}$. Commas in subscripts are used to distinguish between properties that are defined for multiple sets, e.g., $n_{g,z}$ and $n_{g,w}$ describe the complexity of the representation of $\mathcal{Z}$ and $\mathcal{W}$, respectively. The $n$-dimensional unit hypercube is denoted by $\mathcal{B}_{\infty}^n=\left\{x\in\Rspace^{n}~|~\|x\|_{\infty}\leq1\right\}$. The set of all $n$-dimensional binary vectors is denoted by $\{-1,1\}^{n}$ and the interval set between a lower bound $b_{l}$ and an upper bound $b_{u}$ is denoted by $[b_{l},b_{u}]$. Matrices of all $0$ and $1$ elements are denoted by $\mathbf{0}$ and $\mathbf{1}$, respectively, of appropriate dimension and $\id$ denotes the identity matrix. The concatenation of two column vectors into a single column vector is denoted by $(g_1,\:g_2)=[g_1^T\:g_2^T]^T$.

Given the sets $\mathcal{Z},\mathcal{W}\subset\Rspace^{n},\:\mathcal{Y}\subset\Rspace^{m}$, and matrix $R\in\Rspace^{m\times n}$, the linear mapping of $\mathcal{Z}$ by $R$ is $R\mathcal{Z}=\{Rz~|~z\in\mathcal{Z}\}$, the Minkowski sum of $\mathcal{Z}$ and $\mathcal{W}$ is $\mathcal{Z}\oplus\mathcal{W}=\{z+w~|~z\in\mathcal{Z},\:w\in\mathcal{W}\}$, the generalized intersection of $\mathcal{Z}$ and $\mathcal{Y}$ under $R$ is $\mathcal{Z}\cap_R\mathcal{Y}=\{z\in\mathcal{Z}~|~Rz\in\mathcal{Y}\}$, and the Cartesian product of $\mathcal{Z}$ and $\mathcal{Y}$ is $\mathcal{Z}\times\mathcal{Y}=\{(z,y)|~z\in\mathcal{Z},\:y\in\mathcal{Y}\}$.

\subsubsection{Hybrid Zonotopes}
\label{sec:HybZono}

\begin{defn}\label{def-hybridZono} \cite[Def. 3]{Bird_HybZono}
The set $\mathcal{Z}_h\subset\Rspace^n$ is a \emph{hybrid zonotope} if there exist $G^c\in\Rspace^{n\times n_{g}}$, $G^b\in\Rspace^{n\times n_{b}}$, $c\in\Rspace^{n}$, $A^c\in\Rspace^{n_{c}\times n_{g}}$, $A^b\in\Rspace^{n_{c}\times n_{b}}$, and $b\in\Rspace^{n_c}$ such that {\small
    \begin{equation}\label{def-eqn-hybridZono}
        \mathcal{Z}_h = \left\{ \left[G^c \: G^b\right]\left[\begin{smallmatrix}\xi^c \\ \xi^b \end{smallmatrix}\right]  + c\: \middle| \begin{matrix} \left[\begin{smallmatrix}\xi^c \\ \xi^b \end{smallmatrix}\right]\in \mathcal{B}_\infty^{n_{g}} \times \{-1,1\}^{n_{b}}, \\ \left[A^c \: A^b\right]\left[\begin{smallmatrix}\xi^c \\ \xi^b \end{smallmatrix}\right] = b \end{matrix} \right\}\:.
\end{equation}}
\vskip \baselineskip
\end{defn}

 A hybrid zonotope is the union of $2^{n_b}$ constrained zonotopes corresponding to the possible combinations of binary factors, thus it can efficiently represent nonconvex and disjoint sets. Comparisons to and between preexisting set representations can be found in \cite{Bird_HybZono,BIRDthesis_2022,Althoff_SetPropagationTechniques_2021}. The hybrid zonotope is given in \textit{Hybrid Constrained Generator-representation} and the shorthand notation of $\mathcal{Z}_h=\langle G^c,G^b,c,A^c,A^b,b\rangle\subset\Rspace^n$ is used to denote the set given by \eqref{def-eqn-hybridZono}. Continuous and binary \emph{generators} refer to the columns of $G^c$ and $G^b$, respectively. A hybrid zonotope with no binary generators is a constrained zonotope, $\mathcal{Z}_c=\langle G,c,A,b\rangle\subset\Rspace^n$, and a hybrid zonotope with no binary generators and no constraints is a zonotope, $\mathcal{Z}=\langle G,c\rangle\subset\Rspace^n$. Identities and time complexity of linear mappings, Minkowski sums, generalized intersections, and generalized half-space intersections are reported in \cite[Section 3.2]{Bird_HybZono}. An identity and time complexity for Cartesian products is given in \cite{BIRDthesis_2022}. Methods for removing redundant generators and constraints of a hybrid zonotope were reported in \cite{Bird_HybZono} and further developed in \cite{BIRDthesis_2022}.
 
\subsubsection{Successor Set}
\label{sec:SuccessorIntro}

Consider a class of discrete-time nonlinear dynamics given by
\begin{align}
    x_{k+1} = f(x_{k},u_{k}) \:,
\end{align}%
with state and input constraint sets given by $\mathcal{X}\subset\Rspace^n$ and $\mathcal{U}\subset\Rspace^{n_u}$. The $i^{th}$ row of $f(x_{k},u_{k})$ is a scalar-valued function and denoted by $f_i(x_{k},u_{k})$. Disturbances are omitted for simplicity of exposition, although the results in this paper can be extended to systems with disturbances. 
Because hybrid zonotopes are the set representation of interest in this paper and are inherently bounded, the following assumption regarding the dynamics is made.

\begin{assumption}
For all $(x,u)\in\X\times\U$, $||f(x,u)||<\infty$.
\end{assumption}

\begin{defn}
The \emph{successor set} from $\mathcal{R}_k\subseteq\Rspace^n$ with inputs bounded by $\U_k\subseteq\U$ is given by 
\begin{align}\label{eqn-Suc}
    \suc(\R_{k},\U_k)=\left\{\begin{matrix}
        f(x,u)
        \mid\:
        x\in\mathcal{R}_k,\: u\in\U_k
    \end{matrix}
    \right\}\:.
\end{align}%
\vskip \baselineskip
\end{defn}
The $k^{th}$ forward reachable set, $\R_{k}$, from an initial set $\R_0$ can be found by $k$ recursions of successor sets \eqref{eqn-Suc}.

\subsubsection{Special Ordered Sets}
\label{sec:SOSintro}

Special Ordered Set (SOS) approximations were originally developed to approximate solutions of nonlinear optimization programs by replacing nonlinear functions with piecewise-linear approximations \cite{beale1970_SOS}. In this section, we define an SOS approximation. In Section~\ref{sec:HybZonoSUS}, \textbf{Theorem \ref{thm-SOS2HYBZONO}} provides an identity to represent an SOS approximation as a hybrid zonotope.

\begin{defn}
An SOS approximation $\Ss$ of a scalar-valued function $f(x)$ is defined by a vertex matrix $V=[v_1,v_2,...,v_{n_v}]\in\Rspace^{(n+1)\times n_v}$ such that $v_i = (x_i,f(x_i))$ and is given by $\Ss=\{ V \lambda\ |\ \mathbf{1}^T \lambda = 1,\ \mathbf{0}\leq\lambda,$ where at most $n+1$ entries of $\lambda\in\Rspace^{n_v}$ are nonzero and correspond to an $n$-dimensional simplex$\}.$
\end{defn}

Partitioning of the domain of $f(x)$ into simplexes is not unique. Delaney triangulation can be used to generate a particular division of $N$ simplexes, which can be represented using an incidence matrix $M\in\Rspace^{n_v \times N}$ with entries of either $0$ or $1$. The $i^{th}$ column $M_{(\cdot,i)}$ corresponds to the $i^{th}$ simplex over the domain, and the corresponding vertices are given by the first $n$ dimensions of $V_{(\cdot,j)}\:\forall\:j $ such that $M_{(j,i)}=1$.
\section{Reachability via State-update Sets}
\label{sect:SUS_reach}

This section first introduces the open-loop state-update set, which encodes all possible state transitions given by $f(\cdot,\cdot)$ over a user-specified domain of states and inputs, and can be used to calculate successor sets of the open-loop system.
Then, after defining a state-input map as all possible inputs of a given control law over a user-specified domain of states, the set of possible state transitions of the closed-loop system is constructed by combining the state-input map and the open-loop state-update set. It is then shown how this closed-loop state-update set can be used to calculate successor sets of the closed-loop system via an algebraic identity.

\begin{defn}
\label{def:FSUS_OL}
The \textit{open-loop state-update set} $\FSUSOL\subseteq\Rspace^{2n+n_u}$ is defined as 
\begin{align}
    \label{eq:ForwardSUS}
    \FSUSOL = \left\{\ \begin{bmatrix}
    x_k\\
    u\\
    x_{k+1}
    \end{bmatrix}\ \bigg |\ \begin{array}{c}
         x_{k+1} \in \suc (\{x_{k}\},\{u\}),\\
         (x_k,u) \in \domfFXU
    \end{array} \right\}\:.
\end{align}
\vskip \baselineskip
\end{defn}%
We refer to $\domfFXU\subset\Rspace^{n+n_u}$ as the \textit{domain set} of $\FSUSOL$, typically chosen as the region of interest for analysis.  

\begin{theorem}
\label{th:OneStep_F_OL}
Given a set of states $\R_k\subseteq\Rspace^n$, a set of inputs $\U_k\subseteq\Rspace^{n_u}$, and an open-loop state-update set $\FSUSOL$, if $\R_{k}\times\U_k\subseteq \domfFXU$, then the \textit{open-loop successor set} is given by
\begin{align}
    \label{eq:1stepForward_OL}
    \suc(\R_k,\U_k) &= \begin{bmatrix}
        \textbf{0} & \id_{n}
        \end{bmatrix}\big(\FSUSOL\cap_{[\id_{n+n_u}~\mathbf{0}]} (\R_{k}\times\U_k)\big) \:.
\end{align}
\begin{proof}
By definition of the generalized intersection,
    \begin{align}
        \nonumber
         \FSUSOL & \cap_{[\id_{n+n_u}~\mathbf{0}]} (\R_{k}\times\U_k)\ \\
        \nonumber & \quad= \left\{ \begin{bmatrix}
    x_k\\
    u\\
    x_{k+1}
    \end{bmatrix}
    \Bigg | \begin{array}{c}
         x_{k+1} \in \suc (\{x_{k}\},\{u\}),\\
         \begin{bmatrix}
         x_k\\
         u
         \end{bmatrix} \in \domfFXU \cap (\R_k \times \U_k)
    \end{array} \right\}\:.
    \end{align}%
    If $\R_{k}\times\U_k\subseteq \domfFXU$, then $\domfFXU \cap (\R_k \times \U_k) = \R_k \times \U_k$, and \eqref{eq:1stepForward_OL} gives
    $
         \{ x_{k+1}  |
         x_{k+1} \in \suc (\{x_{k}\},\{u\}),\: 
         x_k\in\mathcal{R}_k,\: u\in\U_k\}\:.
   $
   \end{proof} 
\end{theorem}
The containment condition in \textbf{Theorem \ref{th:OneStep_F_OL}}, $\R_{k}\times\U_k\subseteq \domfFXU$, is not restrictive as modeled dynamics are often only valid over some region of states and inputs, which the user may specify as $\domfFXU=\X\times\U$.

Consider a set-valued function $\C(x_k)$ corresponding to a state-feedback controller, such that $\C(x_k)$ is the set of all possible inputs that the controller may provide given the current state, $x_k$. For example, the set-valued function of a linear feedback control law $u(x_k)=Kx_k$ is given by the column vector $\C(x_k) = \{Kx_k\}$.
However, in the case of a linear feedback control law with actuator uncertainty $u=Kx_k+\delta_u$, where $\delta_u \in \Delta_u$, results in the set-valued function $\C(x_k) = \{Kx_k+\delta_u\ |\ \delta_u\in\Delta_u\}$. The \textit{state-input map} encodes the feedback control law given by $\C(x_k)$ as a set over a domain of states, $\text{D}(\Theta)$.
\begin{defn} \label{def:StateInputMap}
The \textit{state-input map }is defined as
\begin{equation}
\SIM=\{(x_k,u)~|~u\in\C(x_k),\ x_k\in\domSIM \} \:,
\end{equation}
where $\domSIM$ is the \emph{domain set} of $\SIM$.
\end{defn}

Similar to $\text{D}(\Psi)$, $\text{D}(\Theta)=\X$ can be specified by the user. Next, the closed-loop state-update set under a controller given by $\C(x_k)$ is defined. Then it will be shown how to construct a closed-loop state-update set given an open-loop state-update set and a state-input map.

\begin{defn}
\label{def:FSUS_CL}
 The \textit{closed-loop state-update set} $\FSUSCL\subseteq\Rspace^{2n}$ for a controller given by $\C(x_k)$ is defined as 
\begin{align}
    \label{eq:ForwardSUS_CL}
    \FSUSCL = \left\{ \begin{bmatrix}
    x_k\\
    x_{k+1}
    \end{bmatrix}\ \bigg |\ \begin{array}{c}
         x_{k+1} \in \suc \left(\{x_{k}\},\C(x_k)\right),\\
         x_k \in \domfFX
    \end{array} \right\}\:,
\end{align}
\vskip \baselineskip
\end{defn}%
where $\domfFX\subset\Rspace^{n}$ is the \textit{domain set} of $\FSUSCL$.

\begin{theorem} \label{thm-CLSUSfromOL} Given an open-loop state-update set $\FSUSOL$ and state-input map $\SIM$, the closed-loop state-update set $\FSUSCL$ with $\domfFX=\begin{bmatrix}
\id_n~\mathbf{0}
\end{bmatrix}\left(\domfFXU\cap\SIM\right)$ is given by
\begin{align}
    \label{eqn-CLfromOL}
    \FSUSCL = \begin{bmatrix}
    \id_n & \mathbf{0} & \mathbf{0}\\
    \mathbf{0} & \mathbf{0} & \id_n\\
    \end{bmatrix} \left( \FSUSOL \cap_{\begin{bmatrix}
    \id_{n+n_u}~\mathbf{0}
    \end{bmatrix}} \SIM\right)\:.
\end{align}
\begin{proof}
By definition of the generalized intersection,
\begin{align}
    \nonumber
    \FSUSOL \cap_{\begin{bmatrix}
    \id_{n+n_u}~\mathbf{0}
    \end{bmatrix}} \SIM &=\\
    \nonumber
    \Bigg \{ \begin{bmatrix}
    x_k\\
    u\\
    x_{k+1}
    \end{bmatrix}\ &\bigg |\ \begin{array}{c}
         x_{k+1} \in \suc (\{x_{k}\},\{u\}),\\
         (x_k,u) \in \domfFXU \cap \SIM,\\
         u\in\C(x_k)
    \end{array} \Bigg \}\:.
\end{align}
Thus the right side of \eqref{eqn-CLfromOL} equals
\begin{align}
    \nonumber
    \Bigg \{ \begin{bmatrix}
    x_k\\
    x_{k+1}
    \end{bmatrix}\ &\bigg |\ \begin{array}{c}
         x_{k+1} \in \suc \left(\{x_{k}\},\C(x_k)\right),\\
         x_k \in [\id_n\ \mathbf{0}] \big(\domfFXU \cap \SIM\big)
    \end{array} \Bigg \}\:.
\end{align}
Comparison to Def.~\ref{def:FSUS_CL} completes the proof.
\end{proof}
\end{theorem}

\textbf{Theorem \ref{th:OneStep_F_CL}} provides an identity for the successor set of a closed-loop system with the feedback control law described by the set-valued function $\C(x_k)$. For closed-loop successor sets, the input set argument $\U_k$ is omitted and the successor set is instead denoted by $\suc(\R_k,\C)$.

\begin{theorem} \label{th:OneStep_F_CL} Given a set of states $\R_k\subseteq\Rspace^n$ and closed-loop state-update set $\FSUSCL$, if $\R_{k}\subseteq \domfFX$, then the\textit{ closed-loop successor set} is given by
\begin{align}
    \label{eq:1stepForward_CL}
    \suc(\R_k,\C) &= \begin{bmatrix}
        \textbf{0} & \id_{n}
        \end{bmatrix}\big(\FSUSCL\cap_{[\id_{n}~\mathbf{0}]} \R_{k}\big) \:.
\end{align}
\begin{proof}
\revres{By definition of the generalized intersection,
{
    \begin{align}
        \nonumber
        \FSUSCL\cap_{[\id_{n}~\mathbf{0}]} \R_{k}\ = \left\{ \begin{bmatrix}
    x_k\\
    x_{k+1}
    \end{bmatrix}
    \Bigg | \begin{array}{c}
         x_{k+1} \in \suc \big(\{x_{k}\},\C(x_k) \big),\\
         x_k \in \domfFX \cap \R_k
    \end{array} \right\}\:.
    \end{align}}%
    If $\R_{k}\subseteq \domfFX$ then $\domfFX \cap \R_k = \R_k$, and \eqref{eq:1stepForward_CL} gives
    $
         \{ x_{k+1}  |
         x_{k+1} \in \suc \left(\{x_{k}\},\C(x_k)\right),\: 
         x_k\in\mathcal{R}_k\}\:.
   $}
   \end{proof} 
\end{theorem}

A fundamental challenge of reachability analysis is that efficient computation of \emph{exact successor} sets is only currently possible for some system classes \cite{Gan_RA4SolvableDynSys}. To obtain formal guarantees for other classes, \emph{over-approximations} of successor sets are often computed instead \cite{WETZLINGER_AdapReachNonLin}. To this end, the following corollaries extend the previous results to over-approximations of successor sets.

\begin{corollary} \label{co-SusOA_ReachOA} For the identities provided by \textbf{Theorems} \textbf{\ref{th:OneStep_F_OL}-\ref{th:OneStep_F_CL}}, if any argument set in the right side is replaced with an over-approximation (e.g., if an over-approximation of the open-loop state-update set, given by $\bar{\FSUSOL}$, is used in place of $\FSUSOL$ in \eqref{eq:1stepForward_OL}), then the identity will instead yield an over-approximation of the left side ($ \suc(\R_k,\U)$).

\begin{proof}
    Set containment is preserved under linear transformation and generalized intersection.
\end{proof} 
\end{corollary}

\begin{remark}
\label{rm:DesiredSuitableOperations} The results of this section are agnostic to set representation, with the exception that the chosen representation must be closed under linear transformation, generalized intersection, and Cartesian product. The reader is directed to \cite[Table 1]{Althoff_SetPropagationTechniques_2021} for a catalog of set representations.
\end{remark}

\section{Reachability of Nonlinear Systems Using Hybrid Zonotopes}
\label{sec:HybZonoSUS}


\begin{assumption} Reachable sets, state-update sets, and state-input maps are represented as hybrid zonotopes.
\end{assumption}

Hybrid zonotopes are closed under linear transformations, generalized intersections \cite{Bird_HybZono}, and Cartesian products \cite{BIRDthesis_2022}. Time complexity of the successor set operations in \eqref{eq:1stepForward_OL} and \eqref{eq:1stepForward_CL} is $\mathcal{O}(n)$, as the linear mappings $[\id\ \mathbf{0}]$ under the generalized intersections amount to matrix concatenations. Set complexity growth of the open-loop and closed-loop successor sets is given by
{\begin{align}
\footnotesize
        \nonumber
        \begin{split}
        \text{\underline{Open}}\\
        n_{g,\text{Suc}} &= n_{g,r}+n_{g,u}+n_{g,\psi}\:,\\
         \nonumber
        n_{b,\text{Suc}} &= n_{b,r}+n_{b,u}+n_{b,\psi}\:,\\
         \nonumber
        n_{c,\text{Suc}} &= n_{c,r}+n_{c,u}+n_{c,\psi}+n\:,
        \end{split}
        \footnotesize
        \nonumber
        \begin{split}
        \text{\underline{Closed}}\\
        n_{g,\text{Suc}} &= n_{g,r}+n_{g,\phi}\:,\\
         \nonumber
        n_{b,\text{Suc}} &= n_{b,r}+n_{b,\phi}\:,\\
         \nonumber
        n_{c,\text{Suc}} &= n_{c,r}+n_{c,\phi}+n\:.
        \end{split}
\end{align}}%
Therefore, iterative calculation of open-loop successor sets using \eqref{eq:1stepForward_OL} results in linear complexity growth dependent on the complexity of $\FSUSOL$ and the same is true of iteration over closed-loop successor sets using \eqref{eq:1stepForward_CL} regarding the complexity of $\FSUSCL$.

The remainder of this section provides a method to represent an SOS approximation of a scalar-valued nonlinear function as a hybrid zonotope and provides an example for $\sin(x)$.

\begin{theorem}
\label{thm-SOS2HYBZONO}
Consider an SOS approximation $\Ss$ defined by the vertex matrix $V=[v_1,\dots,v_{n_v}]\in\Rspace^{(n+1)\times n_v}$ and the incidence matrix $M\in\Rspace^{n_v \times N}$ corresponding to $N$ simplexes, with entries $M_{(j,i)}\in\{0,1\} \:\forall \:i,j$, such that the $i^{th}$ simplex of the partitioned domain is given by the first $n$ dimensions of $V_{(\cdot,j)}\:\forall\:j\in\{k \ |\ M_{(k,i)}=1\}$. Define the hybrid zonotope
\begin{equation}\nonumber
    \mathcal{Q}=\frac{1}{2}\left\langle\begin{bmatrix}
        \mathbf{I}_{n_v} \\ \mathbf{0}
    \end{bmatrix},\begin{bmatrix}
        \mathbf{0} \\ \mathbf{I}_{N}
    \end{bmatrix},\begin{bmatrix}
        \mathbf{1}_{n_v} \\ \mathbf{1}_{N}
    \end{bmatrix}, \begin{bmatrix}
        \mathbf{1}_{n_v}^T\\ \mathbf{0}
    \end{bmatrix}, \begin{bmatrix}
        \mathbf{0}\\ \mathbf{1}_{N}^T
    \end{bmatrix},\begin{bmatrix}
        2-n_v \\ 2 -N
    \end{bmatrix}\right\rangle\:,
\end{equation}
the polyhedron $\mathcal{H}=\{h\in\Rspace^{n_v}~\vert~ h\leq\mathbf{0}\}$, and let
\begin{equation}\label{prop-SOS-eqn-basis}
    \mathcal{D}=\mathcal{Q}\cap_{[\mathbf{I}_{n_v}~-M]}\mathcal{H}\:.
\end{equation}
Then the SOS approximation $\Ss$ is equivalently given by the hybrid zonotope
\begin{equation}\label{prop-SOS-eqn}
    \Z_{SOS}=\begin{bmatrix}
        V &\mathbf{0}
    \end{bmatrix}\mathcal{D}\:.
\end{equation}
\begin{proof}
    Let $\mathcal{D}$ be the hybrid zonotope given by \eqref{prop-SOS-eqn-basis}. 
    For any $(\lambda,\delta)\in\mathcal{D}$ there exists some $(\xi^c,\xi^b)\in\mathcal{B}_{\infty}^{n_v}\times\{-1,1\}^{N}$ such that $\mathbf{1}^T_{n_v}\xi^c=2-n_v$, $\mathbf{1}^T_{N}\xi^b=2-N$, $\lambda=0.5\xi^c+0.5\hspace{1pt}\mathbf{1}_{n_v}$, $\delta=0.5\xi^b+0.5\hspace{1pt}\mathbf{1}_{N}$, and $\lambda-M\delta\in\mathcal{H}\implies\lambda\leq M\delta$. 
    Thus $\lambda\in[0,1]^{n_v}$, $\delta\in\{0,1\}^{N}$, $\sum_{i=1}^{n_v}\lambda_i=1$, and $\sum_{i=1}^{N}\delta_i=1$ results in $\delta_i=1\implies\delta_{j\not=i}=0$. 
    Let $\delta_i=1$, then $\lambda\leq M\delta$ enforces $\lambda_j\in[0,1]\:\forall\:j\in\{0,\dots,n_v\}$ such that $M_{(j,i)}=1$ and $\lambda_k=0\:\forall\:k\not=j$. Therefore given any $z\in\mathcal{Z}_{SOS}$ and $\delta_i=1$, $z=\sum \lambda_j v_j\:\forall\:j\in\{0,\dots,n_v\}$ such that $M_{(j,i)}=1$, thus $ z\in~\Ss$ and $\mathcal{Z}_{SOS}\subseteq~\Ss$. 
    
    Conversely, given any $x\in\Ss$, there exist at most $n+1$ non-negative scalars $\lambda_j\in[0,1]$ corresponding to the $i^{th}$ simplex of the partitioned domain such that $\sum_{j=1}^{n_v}\lambda_j =1$ and $x=\sum\lambda_jv_j$. Let $\delta_i=1$ corresponding to the $i^{th}$ simplex defined by $M_{(\cdot,i)}$ and $\delta_{j\not=i}=0$, then $\lambda\leq M\delta$. Again let $\lambda=0.5\xi^c+0.5\hspace{1pt}\mathbf{1}_{n_v}$ and $\delta=0.5\xi^b+0.5\hspace{1pt}\mathbf{1}_{N}$, thus $\mathbf{1}^T_{n_v}\xi^c=2-n_v$, $\mathbf{1}^T_N\xi^b=2-N$, and $(\xi^c,\xi^b)\in\mathcal{B}_{\infty}^{n_v}\times\{-1,1\}^{N}$. Therefore $(\lambda,\delta)\in\mathcal{D}$, $x=[V~\mathbf{0}](\lambda,\delta)\in\mathcal{Z}_{SOS}$, $\Ss~\subseteq\mathcal{Z}_{SOS}$, and $\mathcal{Z}_{SOS}=\Ss$. 
\end{proof}
\end{theorem}

\begin{figure*}[htb!]
    \centering
    \includegraphics[width=0.75\linewidth]{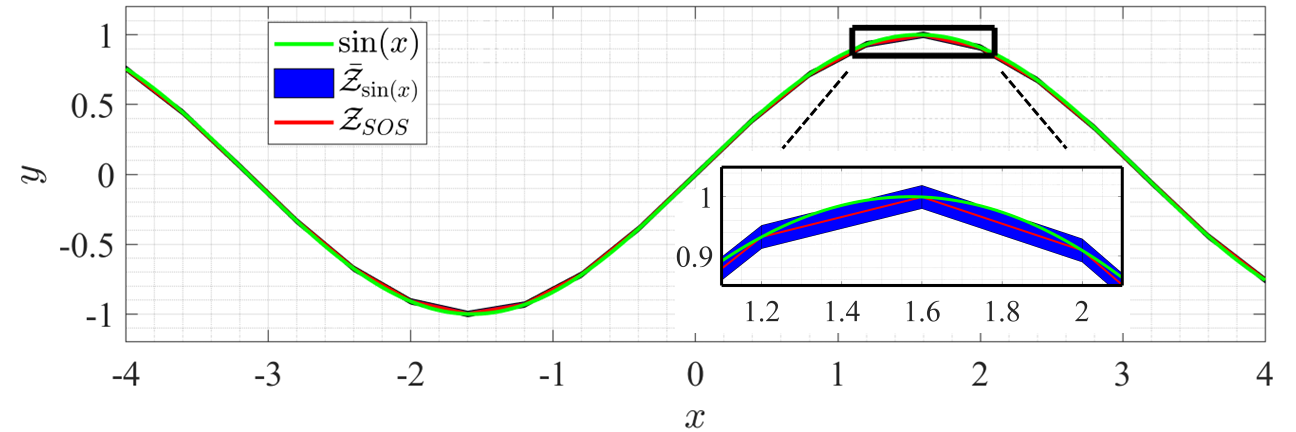}
    \caption{A sinusoid (green) is approximated using an SOS approximation for $x\in[-4,4]$ and represented as a hybrid zonotope (red). Using formal bounds for SOS approximation error, the SOS approximation is bloated to create an enclosure of a sine wave for $x\in[-4,4]$, which is also represented as a hybrid zonotope (blue).}
    \label{fig-SINXHZ}
\end{figure*}

\begin{ex} \label{ex-sinxHZ}
Figure \ref{fig-SINXHZ} shows $y=\sin(x)$ (green) for $x\in[-4,4]$ and an SOS approximation (red) with vertex matrix $V\in\Rspace^{2\times21}$
{\small
\nonumber
\begin{align}
    V = \begin{bmatrix}
    -4 & -3.6 & -3.2 & \dots & 4\\ 
    \sin(-4) & \sin(-3.6) & \sin(-3.2) & \dots & \sin(4) 
    \end{bmatrix} \:,
\end{align}}%
and incidence matrix $M\subset\Rspace^{21\times20}$
{\small
\begin{align}
\nonumber
    M = \begin{bmatrix}
        1 & \mathbf{0} & \mathbf{0}\\
        1 & \ddots & \mathbf{0}\\
        \mathbf{0} & \ddots & 1\\
        \mathbf{0} & \mathbf{0} & 1
    \end{bmatrix} \:.
\end{align}}
The SOS approximation $\Z_{SOS}$ is represented as a hybrid zonotope using \textbf{Theorem \ref{thm-SOS2HYBZONO}}. An envelope $\Bar{\Z}_{\sin(x)}$ (blue) of $\sin(x)$ on $x\in[-4,4]$ is calculated by
\begin{align}
    \nonumber 
    \mathcal{E}_{SOS} &= \langle (0,\delta_{SOS}),\textbf{0} \rangle\:,\\
    \nonumber
    \Bar{\Z}_{\sin(x)} &= \Z_{SOS} \oplus \mathcal{E}_{SOS}\:,
\end{align}%
such that $\Bar{\Z}_{\sin(x)} \supset \left\{ (x,sin(x))\ |\ x\in[-4,4] \right\}$.
The scalar error bound $\delta_{SOS}$ for the SOS approximation of $\sin(x)$ is provided by \cite[Chapter 3]{wanufelle2007global}, along with error bounds for SOS approximations of a variety of other nonlinear functions. 
\end{ex}

Given \textbf{Theorem \ref{thm-SOS2HYBZONO}} and rigorous error bounds for SOS approximations, it is possible to create an over-approximation of the state-update set for a nonlinear dynamic system. While a generalized process for doing this falls outside the scope of this paper and is left to future work, we demonstrate the process with the following numerical example.

\section{Numerical Example}
\label{sect:Examples}

Consider the nonlinear discrete-time dynamics given by
\begin{align}
    \label{eqn-PenulumEulerDyn}
    \begin{bmatrix}
    x_{1,k+1}\\
    x_{2,k+1}
    \end{bmatrix} = 
    \begin{bmatrix}
    x_{1,k} + 0.1 x_{2,k}\\
    \sin(x_{1,k}) + x_{2,k} + 0.1 u_k\\
    \end{bmatrix}\:,
\end{align}
in closed loop with the saturated linear feedback control law
\begin{align}
    \label{eqn-LQRSATlaw}
    u_k(x_k) = \begin{cases}
       Kx_k \:, & \text{if } -20<Kx_k<20\:,\\
        -20 \:, & \text{if } Kx_k \leq-20\:,\\
        \phantom{-}20 \:, & \text{if } Kx_k \geq\phantom{-}20\:,\\
    \end{cases}
\end{align}
where $K = [{-}17.60 \: \: {-}5.61]$. Note that $K$ corresponds to the feedback gain of a linear quadratic regulator (LQR) for the linearization of \eqref{eqn-PenulumEulerDyn} about $x_{1,k}=0$, obtained by replacing $\sin(x_{1,k})$ with $x_{1,k}$, with the state and input weights as identity matrices. The second two cases of \eqref{eqn-LQRSATlaw} represent input saturation. The region of interest is chosen as $(x_1,x_2)\in\X = [-4,4]\times[-8,8]$. The bounded input set is given as $\U\in[-20,20]$. The remainder of this numerical example demonstrates the following procedure to calculate over-approximated reachable sets:
\begin{enumerate}
    \item Construct an over-approximation of the open-loop state-update set $\bar{\FSUSOL}\supset\FSUSOL$ as a hybrid zonotope using \textbf{Theorem \ref{thm-SOS2HYBZONO}}.
    \item Construct an over-approximation of the closed-loop state-update set $\bar{\FSUSCL}\supset\FSUSCL$ as a hybrid zonotope using \textbf{Theorem \ref{thm-CLSUSfromOL}} and \textbf{Corollary \ref{co-SusOA_ReachOA}}.
    \item Calculate forward reachable sets using \textbf{Theorem \ref{th:OneStep_F_CL}} and \textbf{Corollary \ref{co-SusOA_ReachOA}}.
\end{enumerate}
We next specify each of these steps for the example system.
\subsubsection{Constructing $\bar{\FSUSOL}$} An over-approximation of the open-loop state-update set $\bar{\FSUSOL}\supset\FSUSOL$ is constructed by
\begin{align}
    \label{eqn-Ex-OLstep1}
    \mathcal{P}_a&= \begin{bmatrix}
    1 & 0 & 0\\
    0 & 1 & 0\\
    0 & 0 & 1\\
    1 & 0.1& 0\\
    0 & 0 & 0
    \end{bmatrix} \big( \X\times\U \big) \oplus \begin{bmatrix}
    0\\
    0\\
    0\\
    0\\
    1
    \end{bmatrix} \B^1_\infty \:,\\
    \label{eqn-Ex-OLstep2}
    \mathcal{P}_b &= \mathcal{P}_a \cap_{\begin{bmatrix}
    1 & 0 & 0 & 0 & 0\\
    0 & 0 & 0 & 0 & 1
    \end{bmatrix}}\bar{\Z}_{sin(x)}\:,\\
    \label{eqn-Ex-OLstep3}
     \bar{\FSUSOL} &= \begin{bmatrix}
    1 & 0 & 0 & 0 & 0\\
    0 & 1 & 0 & 0 & 0\\
    0 & 0 & 1 & 0 & 0\\
    0 & 0 & 0 & 1 & 0\\
    0 & 1 & 0.1 & 0 & 1
    \end{bmatrix} \mathcal{P}_b\:,
\end{align}%
where $\bar{\Z}_{sin(x)}$ is found using \textbf{Theorem \ref{thm-SOS2HYBZONO}} (see Example~\ref{ex-sinxHZ} and Fig.~\ref{fig-SINXHZ}). Equation \eqref{eqn-Ex-OLstep1} establishes a domain over the region of interest and all possible inputs, and enforces the discrete dynamics of $x_{1,k+1}$ from \eqref{eqn-PenulumEulerDyn} on $p_{a,4}$. Minkowski sum with the interval set in \eqref{eqn-Ex-OLstep1} provides a sufficient basis for $p_{a,5}$ for the generalized intersection in \eqref{eqn-Ex-OLstep2}. Equation \eqref{eqn-Ex-OLstep2} enforces that $(p_{b,1},p_{b,5}) \in \bar{\Z}_{sin(x)}$. The linear transformation in \eqref{eqn-Ex-OLstep3} enforces the remaining terms of the $x_{2,k+1}$ dynamics from \eqref{eqn-PenulumEulerDyn} on $\psi_5$.

\subsubsection{Constructing $\bar{\FSUSCL}$} 
\begin{figure}
    \centering
    \begin{subfigure}[b]{3in}
         \centering
         \includegraphics[width=3in]{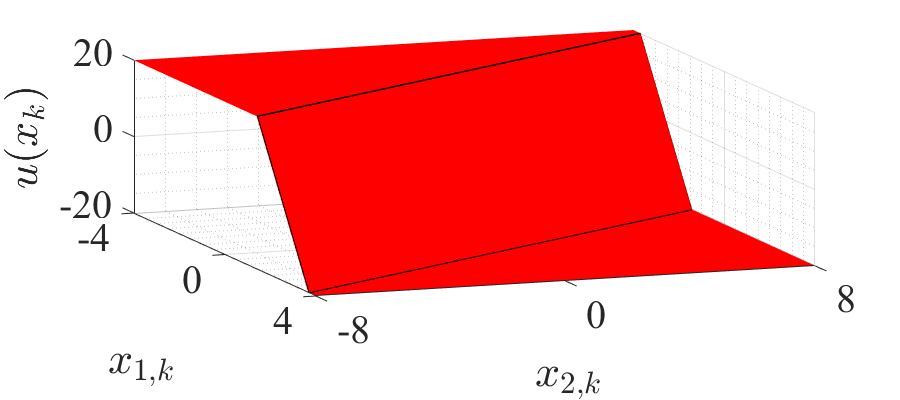}
         \caption{$\SIM$}
         \label{fig:UX_LQRSAT}
     \end{subfigure}\\
    \begin{subfigure}[b]{3in}
         \centering
         \includegraphics[width=3in]{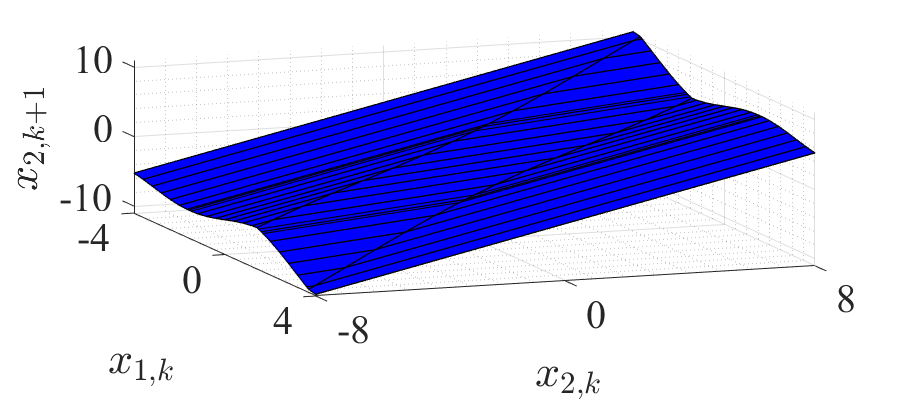}
         \caption{Projection of $\bar{\FSUSCL}$}
         \label{fig:SUS124}
     \end{subfigure}
    \caption{(a)~{State-input map $\SIM$ resulting from \eqref{eqn-LQRSATlaw}}. (b)~Closed-loop nonlinear state-update set for the dynamics given in \eqref{eqn-PenulumEulerDyn} and the state-input map resulting from \eqref{eqn-LQRSATlaw}, projected to dimensions corresponding to $x_{1,k}$, $x_{2,k}$, and $x_{2,k+1}$.}
    \label{fig:SIMandSUS124}
\end{figure}

Equation \eqref{eqn-LQRSATlaw} is a piecewise-affine control law, where each affine control law is defined over a convex region of states. Thus $\SIM$ is first represented over the defined region of interest using a collection of constrained zonotopes. Because hybrid zonotopes are closed under unions \cite{Bird_HybZonoUnionComp}, these constrained zonotopes are then combined to represent the state-input map $\SIM$ as a single hybrid zonotope, shown in Figure \ref{fig:SIMandSUS124}(a). With $\SIM$ and $\bar{\FSUSOL}$ represented as hybrid zonotopes, \textbf{Theorem \ref{thm-CLSUSfromOL}} and \textbf{Corollary~\ref{co-SusOA_ReachOA}} are used to construct an over approximation of the closed-loop state-update set. It can be shown by the construction of $\bar{\FSUSOL}$ and $\SIM$ that $\domfFX=[-4,4]\times[-8,8]$. For visual confirmation, $\bar{\FSUSCL}$ is projected onto dimensions corresponding to $x_{1,k}$, $x_{2,k}$, and $x_{2,k+1}$, as shown in Figure \ref{fig:SIMandSUS124}(b). Notice how partitions associated with the approximation of $\sin(x_k)$ and associated with the piecewise-affine state-input map are recognizable. Furthermore, given any $(x_{1,k},x_{2,k})$, thickness in the $x_{2,k+1}$ direction is difficult to discern, suggesting that $\bar{\FSUSCL}$ appears to be a tight over-approximation of $\FSUSCL$. While a similar projection onto the $x_{1,k}$, $x_{2,k}$, and $x_{1,k+1}$ dimensions is not shown, this would have no over-approximation error in the $x_{1,k+1}$ dimension as their relationship in~\eqref{eqn-PenulumEulerDyn} is linear and therefore represented exactly by $\bar{\FSUSCL}$.

\subsubsection{Forward Reachability} 
\begin{figure}[htb!]
    \centering
    \begin{subfigure}[b]{3.3in}
         \centering
         \includegraphics[width=3in]{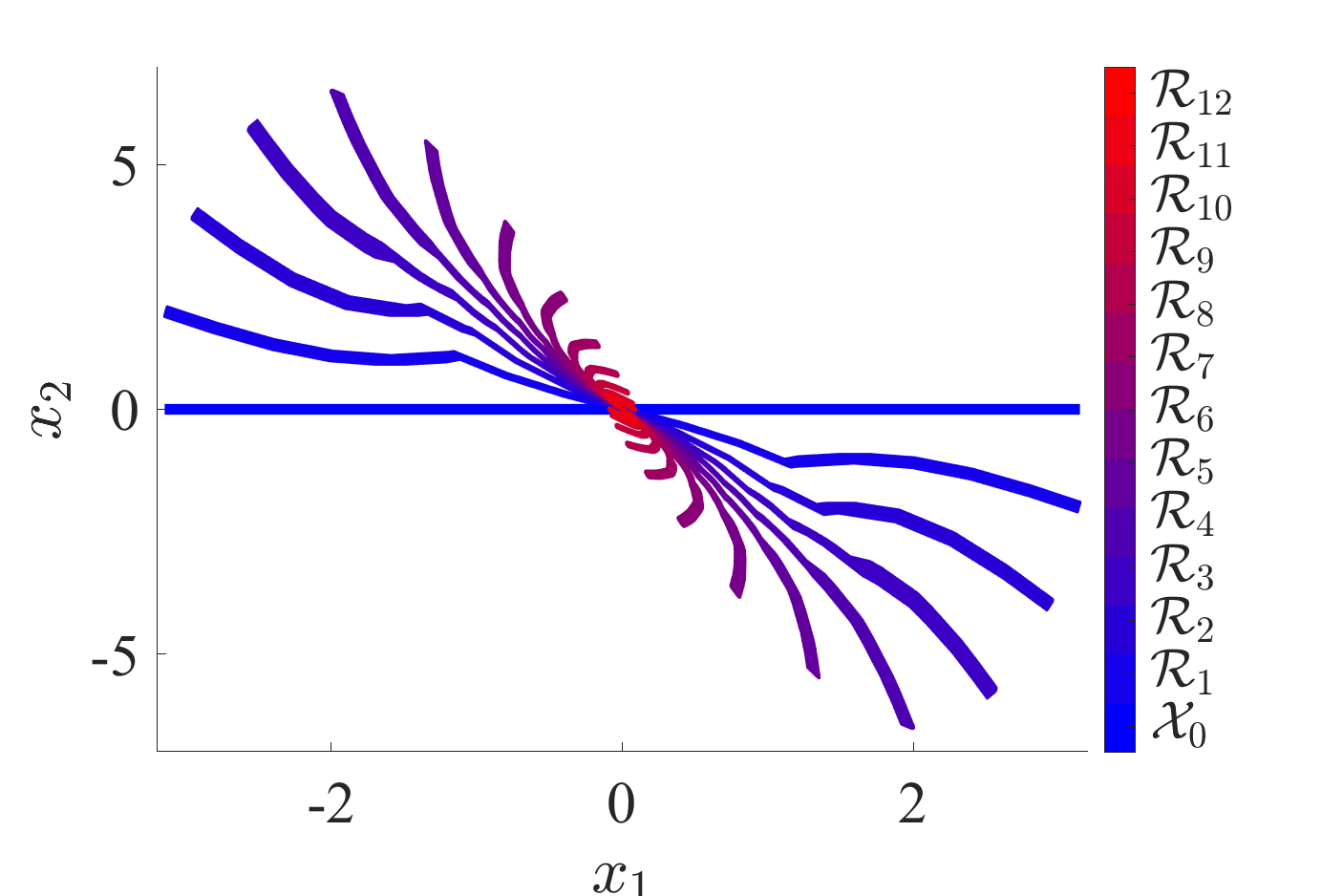}
         \caption{}
         \label{fig:ReachNoTraj}
     \end{subfigure}\\
    \begin{subfigure}[b]{3.3in}
         \centering
         \includegraphics[width=3in]{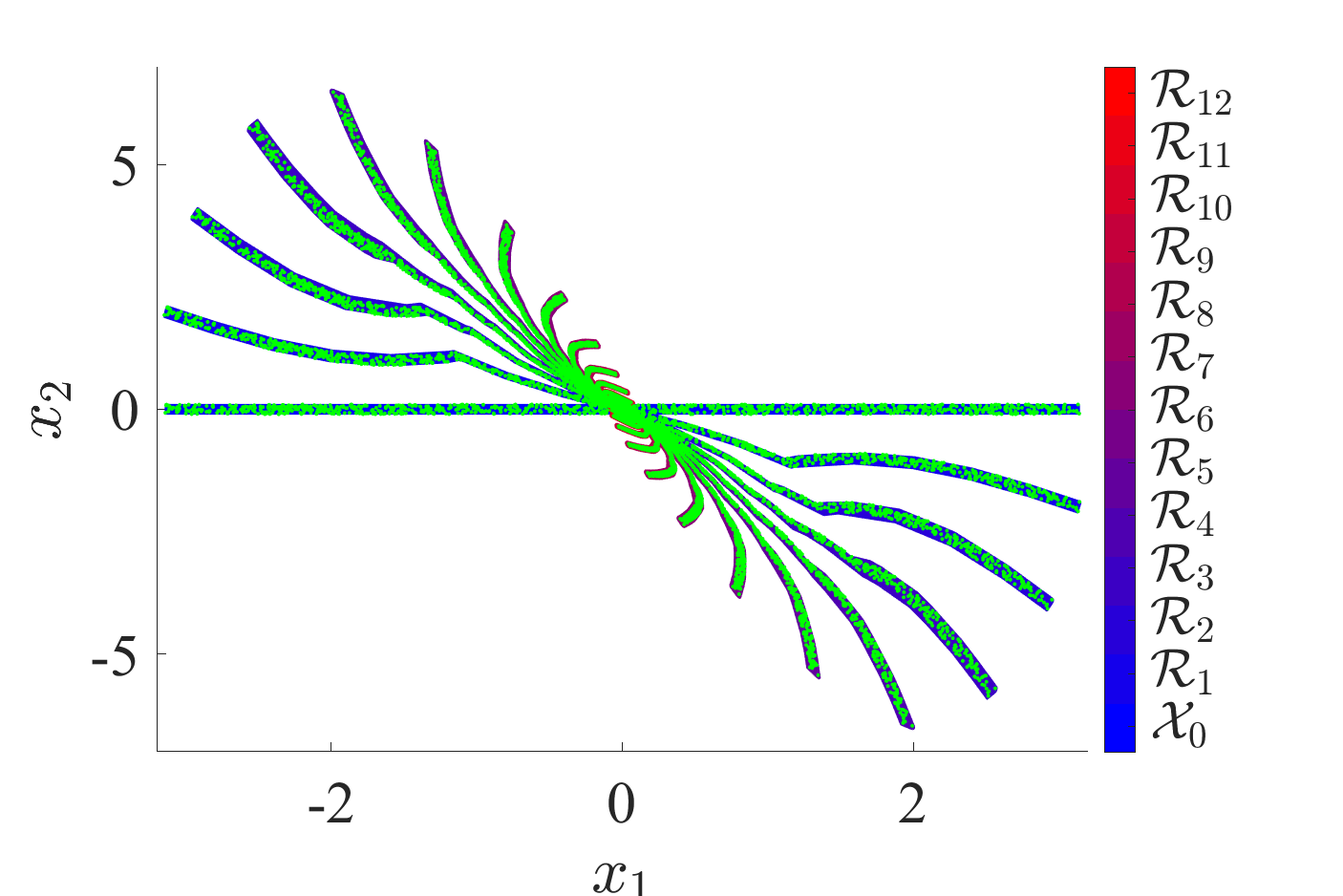}
         \caption{}
         \label{fig:ReachWithTraj}
     \end{subfigure}
     \begin{subfigure}[b]{3.3in}
        \centering
        \includegraphics[width=3.3in]{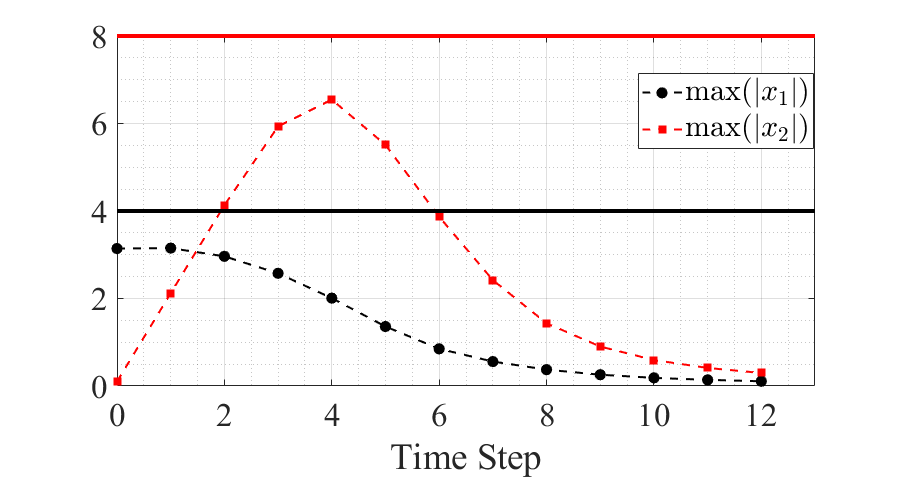}
        \caption{}
        \label{fig:Bounds}
    \end{subfigure}
    \caption{(a) Over-approximation of reachable sets of nonlinear system \eqref{eqn-PenulumEulerDyn} in closed-loop with saturated LQR \eqref{eqn-LQRSATlaw}. (b)~Closed-loop trajectories of the nonlinear system from a random sampling of initial conditions within $\X_0$, plotted over the over-approximations of forward reachable sets. (c)~Maximum magnitudes of $x_1$ and $x_2$ plotted for each time step. Solid lines with matching colors correspond to $\domfFX$ in the $x_1$ and $x_2$ dimensions.}
    \label{fig:Reach}
\end{figure}

Using \textbf{Corollary \ref{co-SusOA_ReachOA}} and iteration over the identity given in \eqref{eq:1stepForward_CL}, over-approximations of forward reachable sets $\R_k$,  $k\in\{1,...,12\}$ are calculated from an initial set given by
\begin{align}
    \X_0 = \left\langle \begin{bmatrix}
    \pi & 0\\
    0 & 0.1
    \end{bmatrix}, \begin{bmatrix}
    0\\
    0
    \end{bmatrix} \right\rangle\:.
\end{align}%
The over-approximated reachable sets are plotted in Fig. \ref{fig:Reach}(a). Figure \ref{fig:Reach}(b) overlays exact closed-loop trajectories, found by randomly sampling $\X_0$ and propagating through the discrete-time nonlinear dynamics. Figures~\ref{fig:Reach}(a)~and~\ref{fig:Reach}(b) exemplify both the nonlinear behavior of the open-loop system and the piecewise-affine behavior of the saturated LQR feedback law. Examination of the exact trajectories suggests that a successful over-approximation of the reachable sets is achieved with relatively small  over-approximation error. Figure \ref{fig:Reach}(c) plots the the maximum magnitude of reachable sets in the $x_1$ and $x_2$ directions, found by solving mixed-integer linear programs, and verifies that the containment condition of \textbf{Theorem \ref{th:OneStep_F_CL}}, $\R_i\subseteq\domfFX$, is met at each time step. These results exemplify how the proposed methods can be used to verify important properties of closed-loop systems, such as satisfaction of safety constraints on states. 

\begin{figure}[htb!]
    \centering
    \begin{subfigure}[b]{3.3in}
        \centering
        \includegraphics[width=3.3in]{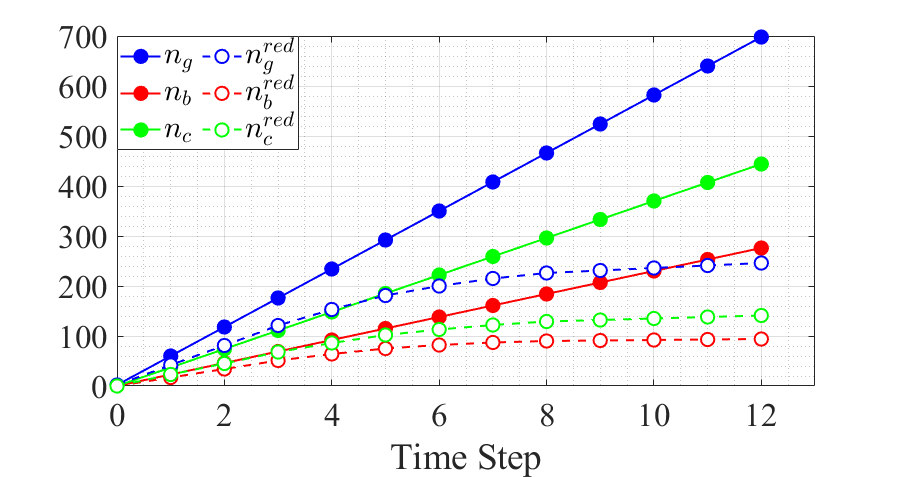}
        \caption{}
        \label{fig:HZComplexity}
     \end{subfigure}\\
     \centering
    \begin{subfigure}[b]{3.3in}
        \centering
        \includegraphics[width=3.3in]{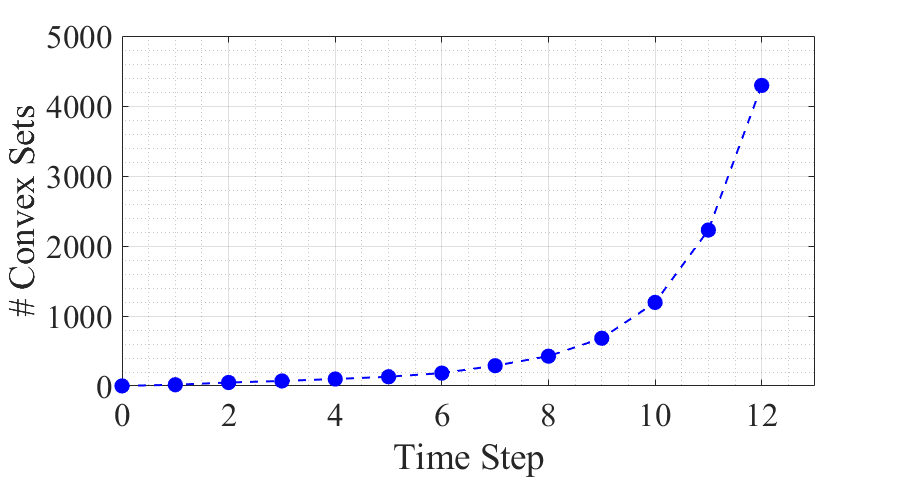}
        \caption{}
        \label{fig:NumNonEmptSets}
    \end{subfigure}
    \caption{(a) Memory complexities of reachable sets, plotted both with and without using redundancy removal techniques from \cite{Bird_HybZono},\cite{BIRDthesis_2022}. Memory complexity without redundancy removal grows linearly. (b) The number of non-empty convex sets represented implicitly by the hybrid zonotopes demonstrates how the proposed methods capture exponential growth in the number of convex sets with linear memory complexity growth.}
    \label{fig:ExComplexity}
\end{figure}

Figure \ref{fig:ExComplexity}(a) plots the memory complexity of the hybrid zonotope reachable sets with and without order reduction from \cite{BIRDthesis_2022} and Fig. \ref{fig:ExComplexity}(b) plots the number of non-empty convex sets that comprise each hybrid zonotope reachable set. These plots illustrate the linear growth in memory complexity of hybrid zonotope reachable sets while representing an exponentially growing number of convex sets.

Results in this section were generated with MATLAB on a desktop computer with a 3.0 GHz Intel i7 processor and 16 GB of RAM. All reachable sets were calculated in $0.02$ seconds and the bounds shown in Figure \ref{fig:Reach}(c) for all time steps were calculated in 23 seconds using the Gurobi mixed-integer optimizer \cite{gurobi_optimization_gurobi_2021}. Reachable sets were plotted using redundancy removal and plotting techniques from \cite{Bird_HybZono,BIRDthesis_2022}. The calculations for removing redundancy and plotting all reachable sets required approximately 10 minutes. A more detailed study of computation times for analysis of hybrid zonotopes can be found in \cite{Bird_HybZono}.

 \section{Conclusion}
\label{sect:conclusion}

This paper presents new methods for calculating over-approximated successor sets of discrete-time nonlinear systems. Using the hybrid zonotope set representation and an over-approximation of the open-loop and closed-loop state-update set, the proposed approach captures worst-case exponential growth in the number of convex sets required to represent nonconvex reachable sets. This is achieved with \textit{linear} growth in memory complexity. Numerical results demonstrate efficient computation and tight over-approximation of reachable sets for a nonlinear discrete-time system in closed-loop with a piecewise-affine control law.

\addtolength{\textheight}{-4.6cm}

\bibliographystyle{IEEEtran}
\bibliography{bibNew}

\end{document}